\documentclass[11pt,onecolumn,amssymb,nofootinbib]{revtex4}
\usepackage{amsmath, amsthm, amscd, amssymb}
\usepackage{graphicx, braket}
\usepackage{bm}
\usepackage{bbm}
\usepackage{epstopdf}

\begin{document}

\title{\bf  Cosmological constant corrections to the photon sphere and black hole shadow radii} \bigskip

\author{Stephen L. Adler}
\email{adler@ias.edu} \affiliation{Institute for Advanced Study,
Einstein Drive, Princeton, NJ 08540, USA}

\author{K. S. Virbhadra}
\email{shwetket@yahoo.com} \affiliation{Mathematics Department, Drexel University, 33rd and Market Streets, Philadelphia, PA 19104, USA}

\begin{abstract}
We review the equations determining the photon sphere radius and the black hole shadow radius, and calculate the cosmological constant corrections arising when the dark energy action has the usual form,  and when dark energy arises from a Weyl scaling invariant dark energy action. For black hole targets of the Event Horizon Telescope, the corrections are very small.

\end{abstract}

\maketitle

\section{Photon sphere and black hole shadow radii}

When the Event Horizon Telescope observes a black hole, what it sees is the black hole shadow.  The shadow is the lensed image at infinity of the photon sphere \cite{evh}, which in turn is the boundary between photon orbits that spiral out to infinity, and photon orbits that spiral inwards into the hole.   For a general static, spherically symmetric line element written in the form
\begin{equation}\label{gen}
ds^2= B(r)dt^2-A(r)dr^2 -D(r) r^2(d\theta^2 + \sin^2\theta d\phi^2)~~~,
\end{equation}
the photon deflection angle in passing from the closest approach radius $r_0$ to a larger radius $R$ is given by \cite{VNC98}
\begin{equation}\label{phifn}
\phi(r_0)- \phi(R)=\int_{r_0}^R \frac{dr}{r} \left(\frac{A(r)}{D(r)}
\right)^{1/2}\left[\left(\frac{r}{r_0}\right)^2 \left(\frac{ B(r_0)D(r)}{B(r)D(r_0)}\right)-1\right]^{-1/2}~~~.
\end{equation}
 A timelike hypersurface ($r=r_{\rm ph}$) in a static spherically symmetric spacetime is defined  \cite{VE00} as a photon sphere if the photon deflection angle for a light ray with closest distance of approach $r_0  = r_{\rm ph}$ becomes unboundedly large.   This happens when the  integral in Eq. \eqref{phifn} becomes divergent at the lower limit of integration  $r_0$,  because the argument of the square bracket in Eq. \eqref{phifn} develops a double zero.  That is, the photon sphere radius is given by the solution of the equation
\begin{equation}\label{rph}
\frac{d}{dr} \left(\frac{r^2 D(r)}{B(r)}\right)=0~~~,
\end{equation}
which when expanded out gives \cite{vir}
\begin{equation}\label{rph1}
2D(r)B(r)+ r \frac{dD(r)}{dr}B(r)-r \frac{dB(r)}{dr} D(r)=0~~~.
\end{equation}
An alternative definition of photon surface for an arbitrary spacetime was given in \cite{cl}, which essentially states that a photon surface in a general  spacetime is a  timelike hypersurace  such that any null geodesic initially tangent to that hypersurface remains tangent to that hypersurface. This definition gives the same result for a static spherically symmetric spacetime as the definition of a photon sphere  in terms of an unbounded photon deflection angle.\footnote{The photon sphere equation Eq. \eqref{rph1} can also give spacelike hypersurfaces (for example, in the  case of  the  Reissner-Nordstr\"om metric \cite{cl}), but these are excluded by the specification of a timelike hypersurface in the general definition of a photon surface.}

Specializing these equations to the polar line element with $D(r)\equiv 1$,
\begin{equation}\label{polar}
ds^2= B(r)dt^2-A(r)dr^2 - r^2(d\theta^2 + \sin^2\theta d\phi^2)~~~,
\end{equation}
the photon deflection equation becomes \cite{wein}
\begin{equation}\label{phifn1}
\phi(r_0)- \phi(R)=\int_{r_0}^R \frac{dr}{r} A^{1/2}(r)\left[\left(\frac{r}{r_0}\right)^2 \left(\frac{ B(r_0)}{B(r)}\right)-1\right]^{-1/2}~~~,
\end{equation}
and the equation determinng the photon sphere radius $r_{\rm ph}$ becomes
\begin{equation}\label{rph2}
\frac{d}{dr} \left(\frac{r^2}{B(r)}\right)=0~~~.
\end{equation}
When expanded out this gives
\begin{equation}\label{rph3}
2B(r)-r \frac{dB(r)}{dr}=0~~~, 
\end{equation}
which is the same as the expansion of the photon sphere equation as written by 
the Event Horizon Telescope collaboration \cite{evh}, 
\begin{equation}\label{rph4}
r=B(r)^{1/2}\left(\frac{d B(r)^{1/2}}{dr} \right)^{-1}~~~.
\end{equation}
Equations \eqref{rph2}--\eqref{rph4} correspond to the standard textbook calculation of the smallest stable photon orbit.  The relation between the black hole shadow radius $r_{\rm sh}$ and the photon sphere radius $r_{\rm ph}$, as given in \cite{evh}, and based on the calculation of Psaltis \cite{ps}, is
\begin{equation}\label{rsh}
r_{\rm sh}=r_{\rm ph}/B(r_{\rm ph})^{1/2}~~~.
\end{equation}
We use Eqs. \eqref{rph3} and \eqref{rsh} as the starting point for the calculations that follow.

\section{Parameterized form for $B(r)$, and cosmological constant corrections to $r_{\rm ph}$ and $r_{\rm sh}$}

``Dark energy'' is usually assumed to arise from the four-space generally covariant action
\begin{equation}\label{darkold}
S_{\rm dark~energy}=-\frac{\Lambda}{8 \pi G} \int d^4x  ({}^{(4)}g)^{1/2}~~~,
\end{equation}
where $\Lambda$ is the ``cosmological constant'', $G$ is Newton's constant, and ${}^{(4)}g=-\det({g_{\mu\nu})}$.
However, as reviewed in \cite{A1}, in a homogeneous cosmology with  $g_{00}(x)\equiv 1$, this action is indistinguishable from a  ``dark energy''  action constructed from
nonderivative metric components so as to be invariant under the Weyl scaling $g_{\mu\nu}(x) \to \lambda(x) g_{\mu\nu}(x)$, where $\lambda(x)$ is a general scalar function.  This novel dark energy action is given by
\begin{equation}\label{dark}
S_{\rm dark~energy}=-\frac{\Lambda}{8 \pi G} \int d^4x  ({}^{(4)}g)^{1/2}(g_{00})^{-2}~~~,
\end{equation}
and gives rise to new cosmological effects only when there are inhomogeneities.  Unlike the standard action of Eq. \eqref{darkold}, the action of Eq. \eqref{dark} is only three-space general coordinate invariant, as a result of the frame
dependence arising from the presence of $g_{00}$.

The factor $(g_{00})^{-2}$ in Eq. \eqref{dark} would become infinite at an event horizon where $g_{00}$ vanishes, and so
black holes as modified by the action of Eq.  \eqref{dark}
differ from their standard general relativity form.   Detailed calculations \cite{F} show that such deviations from a standard Schwarzschild black hole appear within a distance of order $2 \times 10^{-18} (M/M_\odot)^2$cm from a nominal horizon radius $2M$, with $M$ the black hole mass and $M_\odot$ the solar mass.  These  modified, ``Schwarzschild-like'' black holes have neither an event horizon \cite{F} nor an apparent horizon \cite{A2}, further  consequences of which are discussed in \cite{A1}.  It thus becomes of interest to look for tests that distinguish between the two candidate dark energy actions.

Consequences of the novel dark energy action for various tests of general relativity, including light deflection and lensing, were recently studied in  \cite{SLA2}.  This was facilitated by introducing parameterized forms for $A(r)$ and $B(r)$, based on the large $r$ expansion of the formulas in \cite{F}, as follows,
\begin{align}\label{paramdef1}
A(r)=&1+2M/r-C_A \Lambda r^2 +D_A \Lambda M r+...,\cr
B(r)=&1-2M/r-C_B \Lambda r^2 +D_B \Lambda M r+...,\cr
\end{align}
with Table I giving the numerical values of the parameters $C_A,\,C_B,\,D_A,\,D_B$ for the line elements arising from the conventional and the Weyl scaling invariant choices of dark energy action.\footnote{The first and second lines of Table I have a different status with respect to a series expansion in $\Lambda$.  The first line comes from the
conventional dark mater action of Eq. \eqref{darkold}, which leads to the Schwarzschild-de Sitter metric and is exact in $\Lambda$.  So the result $r_{ph}=3M$  obtained from the first line
is exact to all orders in $\Lambda$.  The second line comes from the order $\Lambda$ term in the power series expansion  that is calculated in \cite{F}.  Terms of order $\Lambda^2$ and higher were not calculated in \cite{F} and are absent from the second line of Table I,
so it is only consistent to calculate the $D_B$ correction to $r_{ph}=3M$  to order $\Lambda$, dropping higher order terms.  That is what is done in Eq. \eqref{soln1}.}

\begin{table} [ht]
\caption{Parameters $C_A,\,C_B,\,D_A,\,D_B$ for the spherically symmetric line element arising from the conventional and the Weyl-scale invariant dark energy actions.}
\centering
\begin{tabular}{c  c c c c c}
\hline\hline
dark energy type&Equation& $~~C_A~~$  & $~~C_B~~$  &$~~D_A~~$  &  $~~D_B~~$  \\
\hline
~~~~~conventional ~~~~~~~~~~& (12)      &  -1/3  & 1/3  & 4/3    &  0  \\
Weyi scaling invariant & (11)      &   1   & 1   &   -10   &  -14 \\
\hline\hline
\end{tabular}
\label{tab1}
\end{table}

Because Eqs. \eqref{rph3} and \eqref{rsh} involve only $B(r)$, the parameterized form of $A(r)$ will not be used in what
follows.  Substituting the parameterized form of $B(r)$ into Eq. \eqref{rph3}, the term $C_B \Lambda$ drops out.\footnote{ Rewriting Eq. \eqref{rph2} as $(d/dr)(B(r)/r^2)=0$, which yields Eq. \eqref{rph3} when expanded out, we see by inspection that since $B(r)/r^2=1/r^2-2M/r^3-C_B \Lambda +D_B \Lambda M /r$, the constant $C_B$ term vanishes when the first derivative is taken.}
So (as noted in \cite{cl} for the conventional dark energy case) there is no order $\Lambda$ correction to the photon sphere radius, which is determined by solving
\begin{equation}\label{soln}
0=-2/r^3+6 M/r^4 - D_B \Lambda M /r^2~~~.
\end{equation}
To leading order in $\Lambda$, neglecting terms of order $\Lambda^2$ and higher, this gives
\begin{equation}\label{soln1}
r_{\rm ph}\simeq 3 M -(1/2) D_B \Lambda M r_{\rm ph}^2  \simeq 3M [1 -(3/2) D_B \Lambda M^2]~~~,
\end{equation}
giving a correction to the photon sphere radius only in the case of the Weyl scaling invariant dark energy action, but not for the conventional action.  For the ratio of the black hole shadow radius to the photon sphere radius, we find from
Eq. \eqref{rsh} that
\begin{equation}\label{ratio}
r_{\rm sh}/r_{\rm ph}=B(r_{\rm ph})^{-1/2}\simeq 3^{1/2}[1- (3 D_B-27 C_B/2 ) \Lambda M^2]~~~.
\end{equation}
In this case, there are corrections of order $\Lambda M^2$ for both the conventional and the Weyl scaling invariant dark energy actions.

The current targets of the Event Horizon Telescope are the M87 black hole, with $M\sim 10^{13}\,{\rm m}$, and the Milky Way black hole SgA*, with $M\sim 10^{10}\,{\rm m}$, both in geometrized units with $G$ and the velocity of light equal to unity.  From $\Lambda \simeq 1.3 \times 10^{-52}\,{\rm m}^{-2}$, these give respectively
$\Lambda M^2 \sim 10^{-26}$ and $\Lambda M^2 \sim 10^{-32}$, so the corrections that we have calculated to $r_{\rm ph}$ and $r_{\rm sh}/r_{\rm ph}$ are extremely small.

\end{document}